\documentclass[aps,amsfonts,pra,twocolumn,showpacs]{revtex4-1}

\usepackage{subfigure}
\usepackage{textcomp}
\usepackage{graphicx}
\usepackage{amssymb}
\usepackage{amsmath}
\usepackage{bm}
\usepackage{amsmath, amsthm, amssymb}
\usepackage{dsfont}
\usepackage[colorlinks=true,linkcolor=blue,urlcolor=blue,citecolor=blue]{hyperref}
\usepackage{multirow}
\usepackage{url}

\DeclareMathOperator{\Tr}{\mbox{Tr}}

\newcommand{\mat}[1]{\mathsf{#1}}

\def\bra#1{\mathinner{\langle{#1}|}}
\def\ket#1{\mathinner{|{#1}\rangle}}

 \global\long\def\av#1{\left\langle #1 \right\rangle }

\newcommand{\rtext}[1]{\textcolor{black}{{#1}}}
\newcommand{\btext}[1]{\textcolor{black}{{#1}}}
\global\long\def\mattowbytwo#1#2#3#4{\left(\begin{array}{cc} #1 & #2 \\ #3 & #4 \end{array}\right)}
\global\long\def\mat#1#2#3#4#5#6#7#8#9{\left(\begin{array}{ccc} #1 & #2 & #3 \\ #4 & #5 & #6\\ #7 & #8 & #9 \\ \end{array}\right)}

\def\beq{\begin{equation}}
\def\eeq{\end{equation}}
\def\bea{\begin{eqnarray}}
\def\eea{\end{eqnarray}}

\begin{document}

\title{Intertwined and vestigial order with ultracold atoms in multiple cavity modes}

\author{Sarang Gopalakrishnan$^{1}$, Yulia E. Shchadilova$^{2}$, and Eugene Demler$^2$}
\affiliation{$^1$Department of Engineering Science and Physics, CUNY College of Staten Island, Staten Island NY 10314 \\ $^2$Department of Physics, Harvard University, Cambridge MA 02138 }

\begin{abstract}
Atoms in transversely pumped optical cavities ``self-organize'' by forming a density wave and emitting superradiantly into the cavity mode(s). 
For a single-mode cavity, the properties of this self-organization transition are well characterized both theoretically and experimentally. Here, we explore the self-organization of a Bose-Einstein condensate in the presence of \emph{two} cavity modes---a system that was recently experimentally realized [Leonard \emph{et al.}, \emph{Nature} {\bf 543}, 87 (2017)].
%
We argue that \rtext{this system can} exhibit a ``vestigially ordered'' phase in which neither cavity mode exhibits superradiance but the cavity modes are mutually phase-locked by the atoms. We argue that this vestigially ordered phase should generically be present in multimode cavity geometries.
\end{abstract}


\maketitle

\section{Introduction}

Strongly correlated condensed-matter systems, such as high-temperature superconductors, are often subject to various distinct ordering tendencies at once. These orders are often nontrivially coupled or \emph{intertwined}, e.g., when superconductivity coexists with charge- and spin-density wave order~\cite{berg_intertwined}. Intertwined order gives rise to hybrid topological defects (such as vortex-dislocation bound states~\cite{agterberg2008, berg_4e, sg_disclin}) as well as to novel, partially melted phases, such as charge-4$e$ superconductors and Ising-nematic phases~\cite{berg_4e, kivelson_VO_pnas}. What is distinctive about these partially melted or ``vestigially ordered'' (VO) states is that their order parameters are \emph{composites} (e.g., products) of the ``parent'' order parameters corresponding to the microscopic ordering tendencies. For example, if the parent orders have order parameters $O_1$ and $O_2$, the vestigial-order phase is one in which $\langle O_1 \rangle = 0$ and $\langle O_2 \rangle = 0$ but $\langle O_1 O_2 \rangle \neq 0$. Although such phases exhibit spontaneous symmetry breaking, they are difficult to identify within mean-field theory, and \rtext{may} arise in solid-state systems because the VO phase is more stable to fluctuations or disorder than the parent phases~\cite{berg_4e, kivelson_VO_pnas, fernandes_VO}.

In the present work, we argue that intertwined and vestigial order arise naturally in systems of ultracold atoms coupled to multiple optical cavity modes (see Fig.~\ref{geometryfig}). We focus on a simple system consisting of a pumped Bose-Einstein condensate confined in two standing-wave cavity modes at a relative angle $\theta$. 
Such a system was recently experimentally realized~\cite{leonard2017supersolid, leonard2}; for two frequency-degenerate cavity modes, a phase was observed with emergent continuous $U(1)$ symmetry breaking (as opposed to the Ising symmetry-breaking in the single-mode problem~\cite{ritschprl, vuletic:prl, baumann, esslinger_ssb, esslinger_softmode, esslinger_structurefactor, esslinger_lattice, hemmerich1, hemmerich2, domokos08, morigi08, nagy:dicke09, piazza2013bose}). This continuous symmetry and the associated Goldstone modes~\cite{leonard2} are indicative of a ``supersolid'' phase~\cite{ritsch_RMP, us}.
The self-organization transition in the single-mode case has been extensively explored; its physical origin is that atoms form a density-wave---which scatters light coherently from the pump laser to the cavity---and at the same time the cavity mode exhibits superradiance~\cite{ritsch_RMP}.

As observed in Refs.~\cite{us, ringcav, leonard2017supersolid}, the analogous transition is richer for multiple degenerate cavity modes, since the system must \emph{select} a mode to superradiate into. We focus here on the two-mode case. Here, the simplest expectation is that there are two intertwined Ising order parameters at the transition, so the symmetry group remains discrete. 
Remarkably, \rtext{there is a regime in which} the ordered state possesses an approximate \emph{emergent} continuous symmetry, which was experimentally observed~\cite{leonard2017supersolid, leonard2}. 
This approximate symmetry emerges when certain nonlinear couplings are \rtext{small}.
In the present work we 
\rtext{explore a different regime of this system} where such nonlinearities are important, and argue that they give rise to a new phase, the ``vestigially ordered'' (VO) phase, which is intermediate between the normal and superradiant phases.
In this phase, the atoms form a density wave but \emph{neither} cavity is superradiant. We identify regimes in which the VO phase can be realized in near-term experiments. We also find that the VO phase is in general \emph{more} stable when the number of relevant cavity modes increases: thus, it is a generic feature of the phase diagram of multimode cavities, although it was missed in previous studies of these systems~\cite{us, ringcav}. 

The physical origin of vestigial order in this context can be summarized as follows. The atomic density wave that forms at the self-organization transition has a short wavelength, corresponding to the momentum difference between the pump and cavity modes. Thus this density wave costs appreciable kinetic energy (on the order of a recoil energy). However, the density wave in the VO phase corresponds to the momentum difference between the two cavity modes, which potentially costs much less recoil energy. Therefore, the VO phase can be energetically cheaper than regular superradiance whenever the difference between these recoil energies is large enough (e.g., if the two cavity modes have similar momenta, or if the kinetic energy is made spatially anisotropic by adding external lattice beams). Moreover, if many cavity modes are involved, the \emph{same} atomic modulation can couple to \emph{multiple} momentum differences between cavity modes. 

The rest of this paper is organized as follows. In Sec.~\ref{model} we write down the general microscopic model we will consider. In Sec.~\ref{polaritons} we review and re-derive the linearized (i.e., Gaussian) theory of self-organization, in terms of atomic and photonic polaritons~\cite{esslinger_softmode, oztop2012, piazza2013bose}. 
In Sec.~\ref{nonlin} we discuss the microscopic origin of the key nonlinear couplings between the polariton modes in the low-energy theory. 
This leads us to a simple (but crucially incomplete) mean-field theory, developed in Sec.~\ref{2modemf}. 
In Sec.~\ref{chimode} we identify low-energy fluctuations that are missed by this mean-field theory, and regimes in which such fluctuations are important.
In Sec.~\ref{main} we develop a low-energy theory including all soft fluctuations, and explore its phase diagram, including the VO phase. 
In Sec.~\ref{expt} we identify experimental parameters for which the VO phase can be detected. 
Sec.~\ref{multimode} extends the analysis to systems with many modes, and argues that the VO phase is stabilized by having multiple cavity modes. Finally, Sec.~\ref{conclusion} summarizes our results and experimental predictions.

\section{Model}\label{model}

We begin with a microscopic model for a Bose-Einstein condensate confined in a standing-wave pump field of the form $\cos(Q_p z)$, interacting with two cavity modes with mode functions $\sim \cos(\mathbf{Q}_i \mathbf{\cdot x})$. We also allow for additional trapping beams along the $z$ direction. We assume that the atomic transition is very far detuned from the optical transitions (as is typically the case in experiments), and that the atoms are high-field seekers, so that all atomic internal states other than the ground state can be adiabatically eliminated. In addition, we work in a frame rotating at the pump laser frequency~\cite{ritsch_RMP}. The Hamiltonian then takes the form

\bea\label{eq1}
\mathcal{H} \!\! & = & \sum_{\mathbf{k}} \frac{k^2}{2M} \psi^\dagger_{\mathbf{k}} \psi_{\mathbf{k}} + \int d^dx V(z) \rho(\mathbf{x}) +  \sum_{n = 1,2} \Delta^0_n a^\dagger_n a_n \\
 & & + \Omega \sum_n \int d^d x (a^\dagger_n + a_n) \rho(\mathbf{x}) \cos(Q_p z) \cos(\mathbf{Q}_n \mathbf{\cdot x}) \nonumber \\ && + \mathcal{G} \sum_{nm} \!\! \int \!\! d^d x (a^\dagger_n a_{m} + \mathrm{h.c.}) \rho(\mathbf{x}) \cos(\mathbf{Q}_n \mathbf{\cdot x})\cos(\mathbf{Q}_{m} \mathbf{\cdot x}) \nonumber
\eea
The coupling constants $\Omega$ and $\mathcal{G}$ can be related to the microscopic parameters for a system of two-level atoms as follows: $\Omega \equiv \eta g / \Delta_A$, and $\mathcal{G} \equiv g^2/\Delta_A$, where $\eta$ is the Rabi frequency of the pump, $g$ is the atom-cavity coupling, and $\Delta_A$ is the detuning of the atomic transition from the pump. However, the form~\eqref{eq1} is general and does not assume that the atoms are two-level. We note that the last line of Eq.~\eqref{eq1} includes terms of the form $\mathcal{G} a^\dagger_n a_n N/2$, where $N$ is the total number of atoms in the cavity. In what follows absorb this constant dispersive shift of the cavity mode due to the atoms into a redefinition of $\Delta_n \equiv \Delta^0_n - \mathcal{G} N/2$. Note also that we have allowed the cavity detunings to be asymmetric, but have assumed symmetric couplings and a symmetric geometry: this case is qualitatively identical to the case where all quantities are asymmetric, so this assumption does not involve any loss of generality.

In realistic experiments there will also be dissipative processes such as photon leakage out of the cavity (at a rate $\kappa$) and atomic spontaneous emission (at a rate $\gamma$). For simplicity, we will assume in the main discussion that these processes are suppressed because the detunings obey $\Delta_n \gg \kappa, \Delta_A \gg \gamma$. We return to the effects of loss in Sec.~\ref{expt}. 

\section{Review of polariton picture}
\label{polaritons}

\begin{figure}[t]
\begin{center}
\includegraphics[width = 0.45\textwidth]{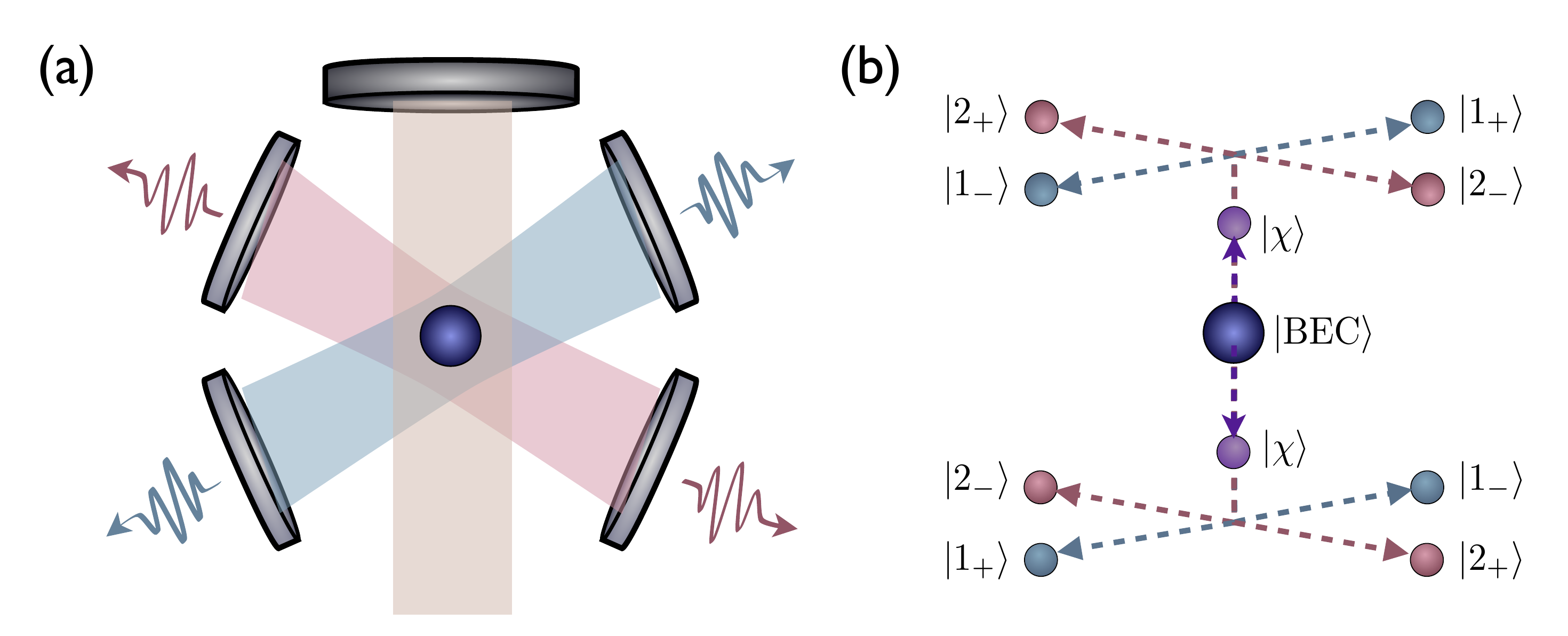}
\caption{Sketch of the geometry considered in this work.  Panel (a) shows the assumed experimental setup, with a retroreflected pump laser and two cavities at an angle $\theta$ relative to one another. The atoms are Bose condensed and lie at the intersection of the two cavity axes and the pump axis. The relevant low-energy atomic momentum modes that are coupled, either directly or indirectly, to light, are shown in the panel (b). 
}
\label{geometryfig}
\end{center}
\end{figure}

We now review the self-organization transition, specializing to the case of Bose condensed atoms with a condensate fraction near unity, and only working to quadratic order in all quantum fields. We discuss the approach to the self-organization transition from the disordered (i.e., non-superradiant) phase, so that the classical expectation values of the cavity modes, i.e., $\langle a_i \rangle = 0$. The atomic fields can be written as $\hat{\psi}_{\mathbf{k}} \simeq \sqrt{N_0} (\delta_{\mathbf{k}, 0} + \mathcal{Q} \delta_{\mathbf{k}, 2Q_p \mathbf{\hat{z}}}) + \hat{\phi}_{\mathbf{k}}$. For simplicity we shall begin with the minimal model that exhibits self-organization; thus we shall neglect both the overall potential term proportional to $V(z)$ and the inter-cavity scattering term in the last line of Eq.~\eqref{eq1}. In the general geometry, the pump and cavity are not perfectly perpendicular (see sketch in Fig.~\ref{geometryfig}). 
Thus, $Q_p \mathbf{\hat{z}} + \mathbf{Q}_n$ and $Q_p \mathbf{\hat{z}} - \mathbf{Q}_n$ are not perfectly degenerate. %
Thus, in general there are two separate atomic Bogoliubov {modes} $\phi_{n\tau} = \sum_{\sigma=\pm} \phi_{\sigma (Q_p \mathbf{\hat{z}} +\tau \mathbf{Q}_n)}$ that couple to the laser. Thus one must in general solve a pair of decoupled three-mode problems, $\lbrace  a_1, \;  \phi_{1,\pm} \rbrace$ and $\lbrace  a_2, \;  \phi_{2,\pm} \rbrace$,

\bea
\mathcal{H}_q & = & \sum_{\sigma=\pm} E_{n\sigma} \phi_{n\sigma}^\dag \phi_{n\sigma} + \Delta_n a^\dagger_n a_n \nonumber \\ & & \quad + \Omega\sqrt{N}(a^\dagger_n + a_n) \sum_{\sigma}(\phi_{n\sigma}^\dag + \phi_{n\sigma} ) \label{quadHam}
\eea
Note that at this level there are no terms coupling the cavities. 

It is convenient to rewrite Eq.~\eqref{quadHam} in terms of coordinate and momentum operators, $\mathbf{\hat q}_n = ( q^{(a)}_n, q^{(\phi)}_{n+}, q^{(\phi)}_{n-})$ and $\mathbf{\hat p}_n = ( p^{(a)}_n, p^{(\phi)}_{n+}, p^{(\phi)}_{n-})$, defined as follows:
\begin{eqnarray}
\hat a_n &=& \sqrt{\frac{\Delta_n}{2}}\left( q_{n}^{(a)} + \frac{i}{\Delta_n} p_{n}^{(a)}\right) \\\ \nonumber
\hat \phi_{n\sigma} &=& \sqrt{\frac{E_{n\sigma}}{2}}\left( q_{n\sigma}^{(\phi)} + \frac{i}{E_{n\sigma}} p_{n\sigma}^{(\phi)}\right) \\\ \nonumber
\end{eqnarray}
In this basis, the Hamiltonian takes the form

\beq
\mathcal{H}_q = \sum_\alpha \left( \frac{p_\alpha^2}{2} + {E_\alpha^2 q_\alpha^2} \right) + \sum_{\alpha \beta}{ q_\alpha Q_{\alpha\beta} q_\beta,} \label{newbasis}
\eeq
where the indices $\alpha,\beta$ run from 1 to 3, indexing the $\mathbf{\hat p}$ and $\mathbf{\hat q}$ vectors introduced above. The main advantage of this basis is that the kinetic energy part (involving the canonical momenta) is proportional to the identity. Thus, to bring this Hamiltonian to diagonal form it suffices to diagonalize the potential-energy terms, which commute with one another. 
{The potential part of the Hamiltonian in the coordinate basis reads}
\begin{equation}
Q_{\alpha\beta,n} =  \frac{\sqrt{N_0}\Omega\Delta_n}{4} 
\mat{\frac{4\Delta_n}{\sqrt{N_0}\Omega}}{\sqrt{ \frac{E_{n+}}{\Delta_n}}  }{\sqrt{\frac{E_{n-}}{\Delta_n}} }
{\sqrt{ \frac{E_{n+}}{\Delta_n}} }{\frac{4 E_{n+}^2}{\Delta_n \sqrt{N_0}\Omega}}{0}
{\sqrt{\frac{E_{n-}}{\Delta_n}} }{0}{\frac{4E_{n-}^2}{\Delta_n\sqrt{N_0}\Omega}} 
 \end{equation} 
We choose the unitary transformation $U_n$ such that it diagonalizes $Q_{\alpha\beta,n} $, $H^{\text{pot}}_n =  \mathbf{\hat q}_{n} Q_n \mathbf{\hat q}_n =  \mathbf{\hat q}_{n} U_n^T \Lambda_{n} U_n \mathbf{\hat q}_n  $, where $\Lambda_n$ is a diagonal matrix of eigenvalues $(\lambda_n)_i, \; i = 1,2,3$. After the unitary transformation we introduce the polariton degrees of freedom $\mathbf{\hat x}_n = U_n \mathbf{\hat q}_n$, $\mathbf{\hat \pi}_n = U_n \mathbf{\hat p}_n$. The commutation relation between new degrees of freedom is preserved, $\left[ \mathbf{\hat x}_n, \mathbf{\hat \pi}_n\right] =\left[ U_n \mathbf{\hat q}_n, \mathbf{\hat p}_n U^T_n \right] = i U_n U^T_n = i I$. 

In the polariton picture the quadratic part of the Hamiltonian is
 \begin{equation}
H_{n}= \sum_\alpha \frac{\hat{\pi}_{\alpha,n}^2}{2} +  \frac{\lambda_{\alpha,n} \hat x_{\alpha,n}^2 }{2}\label{hpol}
\end{equation}
There are correspondingly three eigenvalue branches, which we denote $(\lambda_n)_i, \; i = 1,2,3$ in ascending order of energy. We note two limiting cases: 

(1) When $|E^+_n - E^-_n| \ll \Omega^2 N / \Delta_n$, the energy difference between the two polariton modes is a small perturbation, the lowest polariton has the approximate annihilation operator 
\begin{equation}
\alpha_n \simeq 1/\sqrt{2} (\phi_n^+ + \phi_n^-) + \Omega \sqrt{N}/\Delta_n (a^\dagger_n + a_n)
\end{equation}
while the ``intermediate'' polariton, $\beta_n \simeq \phi_n^+ - \phi_n^-$, softens only weakly. 

(2) When $|E^+_n - E^-_n| \gg \Omega^2 N / \Delta_n$, the two polaritons soften independently. Again, each polariton acquires a small admixture of the cavity mode, but the overall shift of each polariton due to the cavity is weaker:

\beq
\varepsilon_n^\pm \simeq \sqrt{E_n^\pm (E_n^\pm - 4 \Omega^2 N/ \Delta_n)}
\eeq

Except for numerical factors, therefore, the overall structure is the same in this case: the lower-energy polariton softens while the other polariton is still at a relatively high energy $\sim E_n$. This justifies a low-energy description of the phase transition in terms of the $\alpha$-polariton branch for each cavity mode. The photonic admixture in $\alpha_n$ generically scales as $\Omega\sqrt{N}/\Delta_n$, although the prefactor varies (by up to a factor of two) depending on the geometry. In what follows we shall work in this low-energy subspace. 

\section{Microscopic origin of nonlinearities}\label{nonlin}

We now incorporate nonlinear couplings into the Hamiltonian~\eqref{hpol}. Even in the single-mode case, such nonlinear couplings are necessary to stabilize the self-organized crystalline phase. In multimode problems, their importance is even greater, as they determine \emph{which} ordered state is selected: at the quadratic level, the system \rtext{may be} equally unstable to self-organizing into any linear combination of modes 1 and 2; nonlinearities are needed to break this degeneracy, and pick out a ``preferred'' basis for self-organization. In the regime where these nonlinearities preserve the degeneracy, an approximate $U(1)$ symmetry exists, as was seen experimentally~\cite{leonard2017supersolid}.

Since nonlinearities are naturally expressed in terms of the microscopic fields, we must re-express them in terms of the polariton modes to construct an effective low-energy theory. We proceed as follows: first, we rotate the coordinates and momenta according to the transformation above; second, we project out terms that do not act on the low-energy subspace. 
Mapping the original Hamiltonian into the low-energy polariton subspace is implemented by using the elements of the inverse unitary transformation $U_n$ such that $( q^{(a)}_n, q^{(\phi)}_{n+}, q^{(\phi)}_{n-})= ( u^{(a)}_n x_n , u^{(\phi)}_{n+} x_n, u^{(\phi)}_{n-} x_n)$ (and the same for $\mathbf{\pi}_n$) where $(U_n^T)_1 = ( u^{(a)}_n  , u^{(\phi)}_{n+} , u^{(\phi)}_{n-} )$.


Before turning to the low-energy theory, we discuss the microscopic origins of the main cubic and quartic nonlinearities. 

\emph{Cubic terms}. Cubic nonlinearities in Eq.~\eqref{eq1} come from terms involving one condensate (either atomic or photonic) and three quantum fields. Cubic nonlinearities are generated in three different ways: (i)~a photon can get Bragg-scattered from one cavity mode to another, giving rise to a term of the form $\mathcal{G} \sqrt{N_0} a^\dagger_1 a_2 (\phi^\dagger_{\mathbf{Q_1 \pm Q_2}} + \mathrm{h.c.})$, (ii)~an atom starting in mode 1 (i.e., in a density-wave of wavevector $\mathbf{Q_1}$) can scatter a photon from the laser into mode 2, getting a momentum kick $\pm \mathbf{Q_2}$, leading to a cubic coupling of the form $\Omega (a_2^\dagger + a_2) \phi^\dagger_{\mathbf{Q_1 \pm Q_2}} \phi_{\mathbf{Q_1}} + \mathrm{h.c.}$; and (iii)~there are cubic interactions among the $\psi$ modes, due to the contact interaction between the atoms, which we shall neglect in what follows. 

Note that owing to quasimomentum conservation, all cubic nonlinearities necessarily involve at least one mode that is outside the polariton subspace (because it lies in the wrong momentum sector). Thus, in the simplest low-energy theory (which involves projection onto a subspace spanned by $\alpha_1$ and $\alpha_2$) these nonlinearities do not appear. A key conclusion of our work is that these nonlinearities in fact have a crucial role in the self-organization transition. 


\emph{Quartic terms}. In a theory consisting purely of the $\alpha_i$ modes, the leading symmetry-allowed nonlinear couplings are quartic in the $\alpha_i$. In practice the dominant nonlinearity stabilizing the ordered state is number conservation: in the self-organized phase, atoms move from the uniform condensate into condensates at wavevectors corresponding to $\phi_n$. Since the total atom number is conserved, the zero-momentum condensate must be depleted as a result. This depletion leads to a weakening of the effective atom-light matrix element, i.e., $\Omega\sqrt{N} \rightarrow \Omega\sqrt{N - \sum\nolimits_n \phi^\dagger_n \phi_n}$.

Weaker quartic terms arise from the contact interaction and terms of the form $\mathcal{G} a^\dagger_1 a_2 \phi^\dagger_1 \phi_2$, as well as from integrating out high-momentum modes that are coupled in by the cubic nonlinearities. These cases are discussed in the Appendix.

\section{Low-energy theory for two cavity modes}~\label{2modemf}

The simplest way to extend the single-cavity analysis to two cavities is to make the assumption that the only relevant low-energy excitations are the two lowest polaritons, $\alpha_n$. Later, we will re-examine the validity of this assumption. At the quadratic level, when the two cavity modes are degenerate, the system can organize into either of these polariton modes or any linear combination of the modes. Nonlinearities will lift this degeneracy and either favor an equal-weight superposition of both cavity modes, or symmetry-breaking between them. When one projects the Hamiltonian onto the subspace spanned by the lowest polaritons $\alpha_i$, the leading such nonlinearities are quartic; this is mandated by the $Z_2$ symmetry of the self-organization transition in each mode.

This class of two-mode problems has been studied extensively; we review the main results. At the level of mean-field theory, we can treat mode occupations as $c$-numbers, $\alpha_i \rightarrow \langle \alpha_i \rangle$. Moreover, we can choose the condensate and pump phases so that $\langle \alpha_i \rangle$ is real. We then arrive at the following generic classical Hamiltonian, which includes all symmetry-allowed terms up to quartic order:

\beq
\mathcal{H}_{2m} = \sum_n r_n \alpha_n^2 + \lambda \left(\sum_n \alpha_n^2\right)^2 + \tilde{\lambda} \alpha_1^2 \alpha_2^2
\eeq
There are two generic possibilities for the phase diagram of $\mathcal{H}_{2m}$, depending on the sign of $\tilde{\lambda}$. When $\tilde{\lambda} > 0$, the system minimizes its energy by breaking the symmetry between the two cavity modes, and condensing entirely into one of the modes. In this case a phase transition happens, for $r_1 = r_2 < 0$, between a phase that is self-organized into mode 1 and one that is self-organized into mode 2. When $\lambda < 0$, a ``mixed'' phase appears in the phase diagram, when $r_1 \sim r_2 < 0$, in which both cavity modes are macroscopically occupied. In the experimental setting~\cite{leonard2017supersolid} the leading nonlinear process can be shown to contribute \emph{exclusively} to $\lambda$: the physical mechanism is that the atom-light coupling (which depends on the number of condensed atoms) is decreased by the depletion of the $k = 0$ condensate when the system self-organizes. Thus, the experimental system approximately realizes a critical phase at $\lambda = 0, r_1 = r_2 < 0$: in this phase, there is an approximate emergent $U(1)$ symmetry corresponding to arbitrary linear combinations of modes 1 and 2. Such an approximate symmetry has been experimentally observed~\cite{leonard2017supersolid}.

We note that this emergent $U(1)$ symmetry is fine-tuned, and is broken by weaker quartic perturbations. However, our focus in the present work is on a more striking feature of this problem, which is that the low-energy theory outlined above is in general inadequate to describe all the low-lying modes that govern the phase transition. 

\begin{figure}[tb]
\begin{center}
\includegraphics[width = 0.45\textwidth]{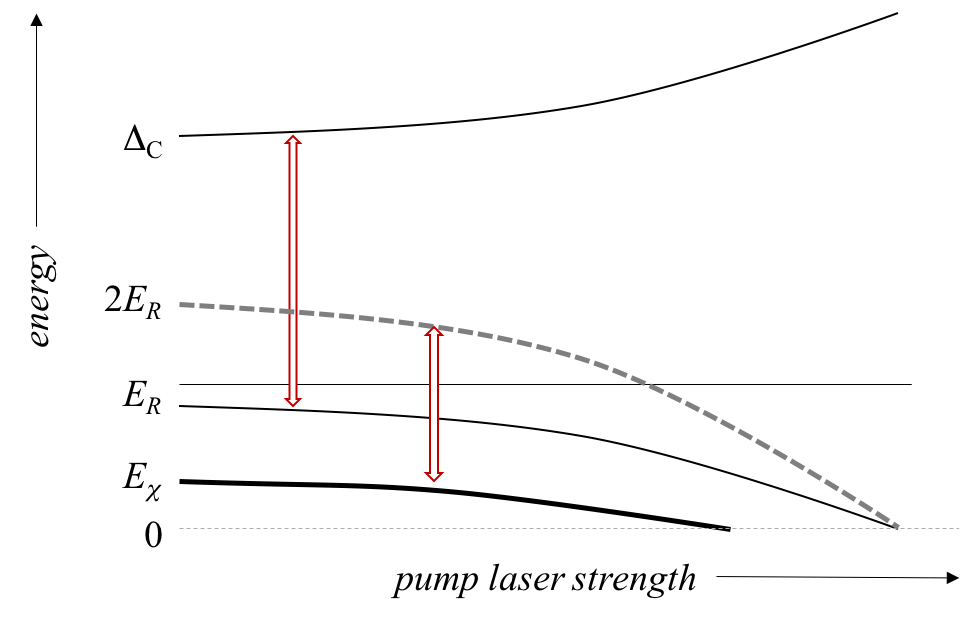}
\caption{Nature of mode-softening in the two-cavity setup. Thin black lines indicate the atomic and photonic polaritons that are familiar from the single-mode problem. These are at energies $E_R$ and $\Delta_C$ in the uncoupled problem, and the former softens giving the usual Dicke transition in a single-mode cavity. Note that there are additional atomic modes (thin gray lines), which are weakly coupled to the photons and are insensitive to the self-organization transition. The $\chi$ mode, which softens to give vestigial order, is depicted by a thick black line. Photon-mediated interactions mix atomic density waves with photons, and also mix the $\chi$ mode with the two-polariton branch (dashed gray line).}
\label{polaritonfig}
\end{center}
\end{figure}

\section{Ancillary soft fluctuations}.~\label{chimode}

In the previous section, we constructed the phase diagram of the theory in which all modes other than the polaritons $\alpha_i$ had been eliminated. We now revisit the validity of this elimination. In general, cubic nonlinearities involve at least one atomic mode that is not directly hybridized with either of the cavity modes. Such atomic modes have momenta $|\mathbf{Q}_1 \pm \mathbf{Q}_2|$. Generically these momenta are of the same order of magnitude as $Q_1, Q_2$, and the corresponding modes are not directly ``softened'' by the atom-light interaction. Thus they usually be adiabatically eliminated, giving rise to weak corrections to the quartic term but no more fundamental changes. 

There are, however, at least two cases in which adiabatic elimination is inappropriate because the \emph{bare} (unsoftened) energy of one such mode is small to begin with. These cases correspond either to (a)~two cavities with a small angle between their axes, or (b)~two cavities at a generic angle, in the presence of a strong pump standing wave. The latter case is straightforward to realize in experiment, as we discuss below.

(a)~When the two cavity axes are at a small angle with respect to one another, the momentum $Q_\chi \equiv |\mathbf{Q}_1 - \mathbf{Q}_2| \approx Q_1 \theta \ll Q_1$. The bare kinetic energy of this mode is $\sim Q_1^2 \theta^2/(2M) \ll E_R$, so it cannot be adiabatically eliminated and must be retained in the low-energy theory. 

(b)~Suppose the two cavity axes are at an arbitrary angle to one another, but are symmetrically arranged with respect to the pump axis (Fig.~\ref{geometryfig}). The difference $\mathbf{Q}_1 - \mathbf{Q}_2$ is parallel to the pump axis, i.e., it points along $\mathbf{\hat z}$. 
We consider a regime in which the pump-induced standing wave is deep, or there is an additional standing wave (of approximately the same wavelength as the pump beam) that lies along the $z$ axis. The single-particle kinetic energy can then be written in the form 

\beq
\epsilon(\mathbf{k}) = \frac{1}{2M} (k_x^2 + k_y^2) + J \cos[k_z \lambda/2],
\eeq
where $\lambda/2$ is the wavelength of the optical lattice, and $J$ is a hopping matrix element along the $z$ axis, which can be computed within a tight-binding model~\cite{greinernote}. 
When the lattice beams are strong, the kinetic energy along $z$ is quenched relative to that along $\mathbf{\hat x}$ and $\mathbf{\hat y}$. Therefore, modes with momenta lying along $\mathbf{\hat z}$ are much lower-energy than those with components along the other axes. Specifically, the kinetic energy of the atomic modes $\phi_{\pm (\mathbf{Q}_1 - \mathbf{Q}_2)}$ can be made arbitrarily small by increasing the lattice strength. 

We emphasize that this mechanism is \emph{not} sensitive to the wavelength of the standing wave along $\mathbf{\hat z}$: the role of this standing wave is simply to suppress the bandwidth of atomic density modes along that direction.

In either of the cases discussed above, the polaritons $\alpha_1$, $\alpha_2$ do not exhaust the low-energy subspace. There is an additional mode, $\chi \equiv \frac{1}{\sqrt{2}} (\phi_{\mathbf{Q}_1 - \mathbf{Q}_2} + \phi_{\mathbf{Q}_2 - \mathbf{Q}_1})$, which must be treated on the same footing. In what follows, we work with this three-mode mean-field theory.

\section{Three-mode low-energy theory} \label{main}

\subsection{Hamiltonian and numerical approach}
\btext{
Including both the cubic and quartic terms from Sec.~\ref{nonlin}, and projecting onto a low-energy subspace spanned by the modes $\alpha_1$, $\alpha_2$, and $\chi$, we arrive at the following effective theory for the three-mode problem:
}
\bea\label{eq:mft}
\mathcal{H}_3 & = & \sum_{n=1,2,\chi} \left( \frac{\hat \pi_n^2}{2} + \frac{\lambda_n \hat x^2_n}{2} \right) + \xi  \hat x_1  \hat x_2 \hat x_\chi  \\
&+& \gamma (\hat x_1^2 +\hat x_2^2) (\hat x_1^2 +\hat x_2^2+\frac{\gamma'}{\gamma}\hat x_\chi^2) \nonumber,
\eea
\btext{where the cubic $\xi$  and quartic non-linearity coefficient $\gamma$, $\gamma'$  are defined using the polariton transformation (see Appendix~\ref{sec:map}). For consistency we rewrite the mode $\chi$ in $\hat x_\chi$ and $\hat \pi_\chi$ with the coefficient $\lambda_\chi=E_\chi^2$. }

We find the ground state of this Hamiltonian numerically, using {the Dirac-Frenkel varitional principle~\cite{Jackiw1979}. This approach is based on a projection of the many-body wave function onto a submanifold of the full Hilbert space spanned by a set of trial wavefunctions in a form of correlated Gaussian state}. We shall turn to these numerical results in Sec.~\ref{sec:gaussian}. To develop intuition for this phase diagram, however, we first discuss its properties in various limits using heuristic arguments. 

\subsection{Three-mode mean-field theory}

We first consider the mean-field theory that one would arrive at by treating all three low-energy fields as classical. Including all symmetry-allowed terms, and taking $\langle \alpha_n\rangle, \langle \chi \rangle$ real, then leads to the phenomenological Hamiltonian


\beq\label{h3mf}
\mathcal{H}_3^{mf}  =  E_1 \alpha_1^2 + E_2 \alpha_2^2 + E_\chi \chi^2 + \xi \alpha_1 \alpha_2 \chi + \ldots
\eeq
where $\ldots$ includes quartic and higher-order terms, which stabilize ordered phases. Note that the cubic term can always be made to lower the energy, by choosing signs appropriately. Thus, the presence of the $\chi$ mode favors a state in which both cavity modes are condensed, as opposed to a state in which only one cavity mode is occupied. Furthermore, the Hamiltonian~\eqref{h3mf} exhibits a first-order phase transition even when all three modes $E_1, E_2, \chi$ are individually stable. The overall structure is shown in Fig.~\ref{pdmf}. 

\subsection{Renormalization effects and vestigial-order phase}\label{renVO}

By treating all fields as $c$-numbers, the mean-field Hamiltonian~\eqref{h3mf} neglects the renormalization of each mode by virtual fluctuations into the others. Consider, in particular, the limit $E_\chi \ll E_1 = E_2$. In this limit, the frequency of the $\chi$ mode is renormalized downward to

\beq
\tilde{E}_\chi = E_\chi - \frac{\xi^2}{E_1 + E_2}.
\eeq
This renormalized mode frequency can go through zero even when all the bare energies are positive, thus triggering a phase transition into a phase in which $\chi$ acquires a macroscopic expectation value but the $\alpha_i$ do not---namely, the vestigial-order (VO) phase.

We now briefly summarize the physical properties of the vestigial-order phase that occurs when the $\chi$ mode softens. In the VO phase, the density wave at momentum $\pm (\mathbf{k}_1 - \mathbf{k}_2)$ is macroscopically occupied. Therefore time-of-flight imaging will show Bragg peaks at this momentum. On the other hand, there is no superradiance in either of the cavity modes. 

By substituting a classical expectation value for the $\chi$ mode (which we, once again, take to be real) into Eq.~\eqref{eq:mft}, we arrive at the following quadratic Hamiltonian for the $\alpha_n$:

\beq\label{hvo}
\mathcal{H}_{VO} = \sum_n \left( \frac{\pi_n^2}{2} + \frac{\lambda_n x_n^2}{2} \right) + \xi \langle \chi \rangle x_1 x_2 + \ldots
\eeq
To solve for the cavity modes in the VO phase, one has to diagonalize Eq.~\eqref{hvo}; the new normal modes of the cavity are symmetric and antisymmetric combinations of the original cavity modes. Whether the symmetric or antisymmetric mode is lower in energy is determined by the sign of $\langle \chi \rangle$: in other words, the degeneracy between the two original cavity modes is spontaneously broken in the VO phase. 
This splitting, and the associated symmetry breaking, can be detected by standard spectroscopic probes of the cavity mode frequency.%

\subsection{Gaussian variational approach and results}
\label{sec:gaussian}


We now extend our previous mean-field theory to account for the renormalization effect that leads to the VO phase. It is simplest to frame this extension in terms of variational wavefunctions. The traditional mean-field theory corresponds to a variational state in which each of the three modes is in a coherent state. We can include the renormalization effect by also allowing the composite order parameters $\langle x_1 x_2 \rangle$, etc., to have an expectation value. In the most general case we describe the system with a Gaussian state 
\begin{equation}\label{eq:WF}
\ket{\Psi} =   e^{-\frac{1}{2}\sum_{ij} v_i \sigma^y_{ij}  \hat R_j - \text{h.c.}} e^{-\frac{i}{4}  \sum_{ij} \hat R_i  Q_{ij} \hat R_j - \text{h.c.}} \ket 0
\end{equation}
where the operator $\hat R = \left( \hat x_1 \ldots \hat x_n  , \hat \pi_1 \ldots \hat \pi_n \right)$ and $n$ running through the polariton modes $\alpha_1$ and $\alpha_2$ as well as the low-energy $\chi$-mode; $\sigma^y$ is an extended Pauli matrix $\sigma^y = \mattowbytwo{0}{-i I_3}{iI_3}{0}$ with an identity matrix $I_3$ of the size three. Here $v_i$ and $Q_{ij}$ are the variational parameters of the state which can be understood as the displacement and squeezing amplitudes correspondingly. \btext{Wavefunction~\eqref{eq:WF} is normalized, since all the exponents are taken to be anti-Hermitian. This is ensured by the equality $Q_{ij}=Q_{ji}$.}

In addition to the naive order parameters, this variational wavefunction allows for squeezing between modes 1 and 2, which is a crucial ingredient in the softening of the $\chi$ mode. For our numerical approach we further allow for squeezing in the other pairs of modes. (Note that \emph{inter-mode} squeezing is the crucial indicator of a phase transition. Squeezing of individual polariton modes is present---i.e., expectation values of the form $\langle \alpha_1 \alpha_1 \rangle \neq 0$---because of Gaussian fluctuations, even in the normal state, as a result of the atom-cavity interaction. However, in the normal state, symmetry dictates that $\langle \alpha_1 \alpha_2 \rangle = \langle \alpha_1 \chi \rangle = \langle \alpha_2 \chi \rangle = 0$.) 

\btext{We minimize the energy using the imaginary time evolution of the parameters of the Gaussian wavefunction, which can be cast in the following form~\cite{kraus2010} 
\begin{equation}\label{eq:DMimtime}
\hat \rho(\tau) = \frac{e^{-\hat H\tau } \hat \rho(0) e^{-\hat H\tau}}{\Tr[ e^{-2\hat H\tau} \hat \rho(0)]}
\end{equation}
From Eq.~\eqref{eq:DMimtime} we derive the differential form for the evolution of the density matrix. We take an infinitesimally small  time step $\tau \rightarrow \Delta \tau$ and expand exponents both in numerator and denominator up to the leading order in $\Delta t$. From this we obtain the following equation for the evolution of the density matrix 
\begin{equation}\label{eq:densitymat}
\partial_\tau \hat \rho(\tau) = - \lbrace \hat H,\hat \rho(\tau) \rbrace + 2 \hat \rho(\tau) \Tr[\hat H \hat \rho (\tau) ]
\end{equation}
Here $\Tr[\hat H \rho (\tau) ]$ is the energy of the system. Averaging one- and two-operators with the density matrix we obtain the equations of motion for the order parameters, ($\av{\hat x}$, $\av{\hat \pi}$)  and two point correlation functions ($\av{\hat x \hat x}$, $\av{\hat \pi \hat x}$ etc.)  in imaginary time. 
We solve those equations numerically for given parameters of the effective low-energy Hamiltonian~\eqref{eq:mft} . The results are shown in Fig.~\ref{pdVO} (and Fig.~\ref{pdgausnovo} in Appendix~\ref{sec:tetracrit}), and include a VO phase for parameters $\xi=1$, $E_\chi=0.01$, and $\gamma=0.1$. }

\btext{
The phase diagram of the system~\eqref{eq:mft} demonstrates five distinct phases (Fig.~\ref{pdVO}). 
In the upper right corner of the phase diagram, all the modes in the low-energy theory are stable, and the system is in a ``trivial'' (or ``normal'') phase with all order parameters equal to zero, $\av{\hat x_1}=\av{\hat x_2}=\av{\hat x_\chi}=0$ (see panels (b-d) in Fig.~\ref{pdVO} of the specific order parameters). Fluctuations $\av{\hat x_n^2}$, however, grow as the phase boundary is approached. When $\lambda_1$ or $\lambda_2$ is decreased, the system undergoes a phase transition from this trivial phase to the superradiant phase in cavity 1 or cavity 2 respectively: e.g., when cavity 1 is superradiant, $\av{\hat x_1}\neq 0, \av{\hat x_2}=\av{\hat x_\chi}=0$. These phases are shown in the upper-left and lower-right corners of Fig.~\ref{pdVO} (a-d). When both parameters $\lambda_1$ and $\lambda_2$ are sufficiently reduced all modes are macroscopically occupied $\av{\hat x_1}\neq 0$, $\av{\hat x_2}\neq0$, and $\av{\hat x_\chi} \neq 0$ (see in the lower left corner in Fig.~\ref{pdVO}(a-d)).  The vestigial order phase,  $\av{\hat x_\chi}\neq 0$ and $\av{\hat x_1}=\av{\hat x_2}=0$ occurs between the trivial phase and the phase in which both modes are occupied, specifically in the center of the phase diagram Fig.~\ref{pdVO} (a-d). }

 
\begin{figure}[tb]
\begin{center}
\includegraphics[width = 0.45\textwidth]{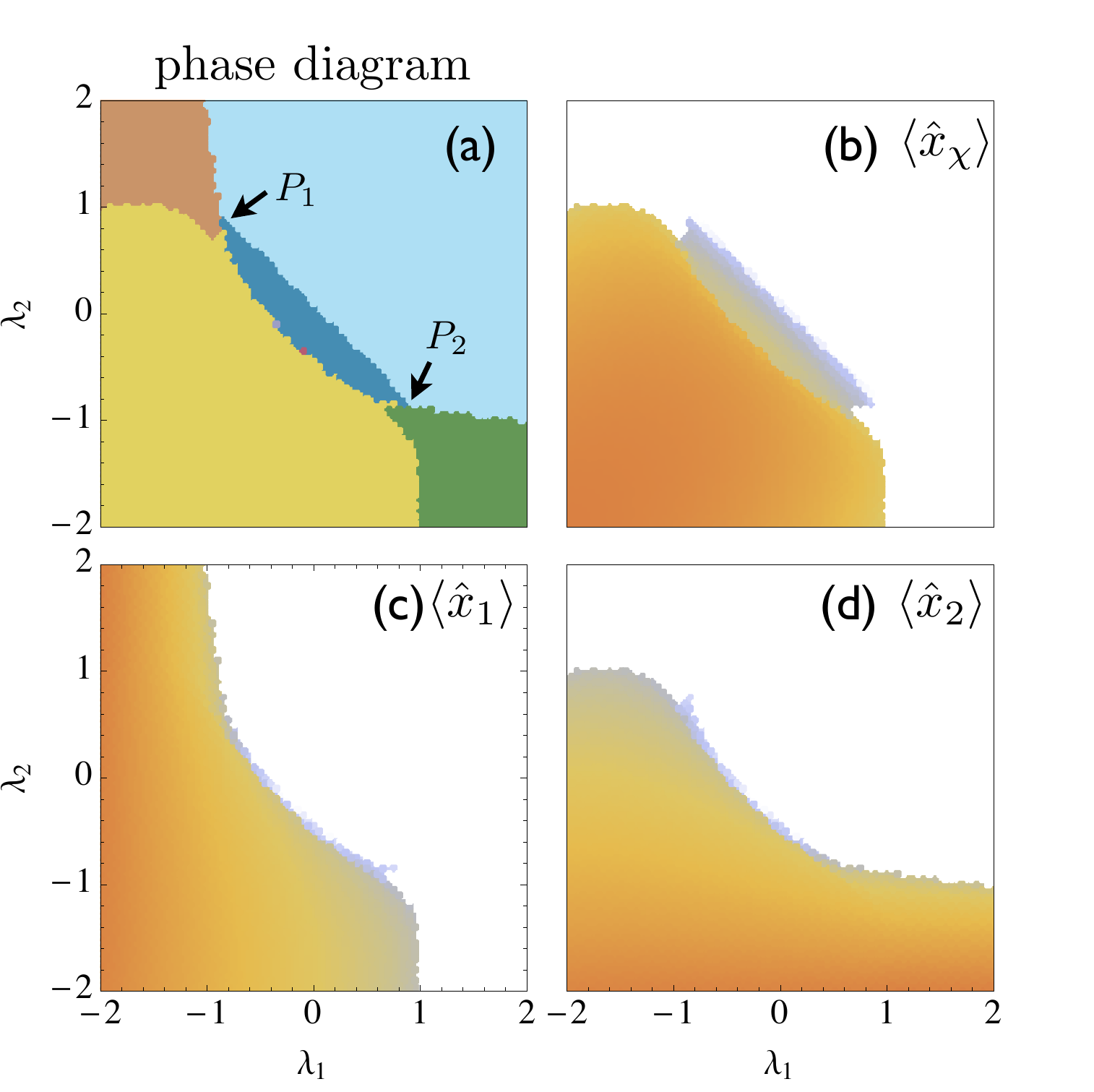}
\caption{(a) Phase diagram of the low-energy Hamiltonian~\eqref{eq:mft} and order parameters (b) $\av{\hat x_\chi}$, (c) $\av{\hat x_1}$, and (d) $\av{\hat x_2}$ as a function of the detunings of the two cavities, $\lambda_1$ and $\lambda_2$. The diagram is calculated for $E_\chi=0.01$, $\xi=1$, and $\gamma=\gamma'=0.1$. Different colors in panel (a) correspond to different phases (see the text for discussion). All phase transitions are of the second order. }
\label{pdVO}
\end{center}
\end{figure}

\subsection{Vestigial phase and multicritical points}

In agreement with our previous, heuristic discussion, we have numerically found that a VO phase exists in the phase diagram of the two-cavity system. As discussed in Sec.~\ref{renVO}, this phase has expectation values $\langle \chi \rangle \neq 0$ as well as $\langle \alpha_1 \alpha_2 \rangle \neq 0, \langle \alpha_1^\dagger \alpha_2 \rangle \neq 0$. These expectation values are zero in the normal phase, and are also zero in the phase where only one cavity is superradiant. However, unlike the phase in which both cavities are ordered, the VO phase has $\langle \alpha_1 \rangle = \langle \alpha_2 \rangle = 0$. The symmetry broken by the VO phase (similar to the superradiant phases) is a discrete Ising symmetry. The VO phase occurs between the trivial phase and the phase with fully broken symmetry. 

The boundary of the VO phase has two multicritical points, $P_1$ and $P_2$, shown in Fig.~\ref{pdVO}(a). Each separates the corresponding individually ordered phase, VO phase, trivial phase, and fully ordered phase. 
For specificity we focus on the vicinity of the point marked $P_1$ in Fig.~\ref{pdVO}(a). Near this point, the order parameter $\langle \alpha_2 \rangle$ is not central to the physics: one can thus write down an effective two-mode description of this region, in terms of the order parameters $\langle \alpha_1\rangle$ and $\langle \chi \rangle \sim \langle \alpha_1 \alpha_2 \rangle$. Since $\langle \alpha_1 \alpha_2 \rangle$ is itself an Ising variable, the resulting two-mode theory takes the same form as in Sec.~\ref{2modemf} for two coupled Ising transitions. There are two possibilities for the phase diagram. 

(a)~The two order parameters compete, such that only one of them is ever present, and there is a first-order phase transition line between a phase with $\langle \alpha_1 \rangle \neq 0, \langle \alpha_1 \alpha_2 \rangle = 0$ and $\langle \alpha_1 \rangle = 0, \langle \alpha_1 \alpha_2 \rangle \neq 0$.  

(b)~The two order parameters cooperate, giving rise to a tetracritical point and a ``mixed'' phase in which both order parameters are present. In this mixed phase, $\langle \alpha_1 \rangle \neq 0, \langle \alpha_1 \alpha_2 \rangle \neq 0$. It immediately follows that $\langle \alpha_2 \rangle \neq 0$. Note that one can think of the mixed phase in two completely equivalent ways: either as featuring condensation of modes 1 and 2, \emph{or} as condensation in one mode plus vestigial order. Although there might seem to be three separate order parameters---$\langle \alpha_1 \rangle, \langle \alpha_2 \rangle, \langle \alpha_1 \alpha_2 \rangle$---there are only two independent Ising symmetries in the problem. The ``third'' order parameter is automatically generated once two Ising symmetries are broken.

Our numerical results indicate that possibility~(b) is what occurs in practice. The normal phase has two separate Ising symmetries. In the mixed phase, both Ising symmetries are broken, so all the order parameters have finite expectation values. The transition between the normal and mixed phases \emph{generically} happens in two stages, with an intervening intermediate phase in which \emph{one} Ising symmetry is spontaneously broken: the order parameter for this phase can be either one of the original Ising variables $\langle \alpha_i \rangle$, or a composite such as $\langle \alpha_1 \alpha_2 \rangle$. At special multicritical points, however, both Ising symmetries can break at once.

\section{Experimental parameters} \label{expt}

We now discuss prospects for observing the predicted phenomena using the cavity parameters of the ETH group. Typically these experiments involve $N \approx 100,000$ particles in a high-finesse single-mode cavity with a linewidth $\kappa \approx 2\pi \times 2$~MHz and a microscopic atom-cavity coupling $g_0 \approx 2\pi \times 10$~MHz. The atomic spontaneous emission linewidth is $\gamma \approx 2\pi \times 3$~MHz. Further, atom loss and heating limit the duration of the experiment to a few seconds; therefore, any experimentally relevant instability needs to have a rate that is at least in excess of $10$~Hz in order to be clearly visible.

The key parameter controlling the VO phase is the coupling $\xi$ in Eq.~\eqref{eq:mft}. In the case of interest this is given by $\sim E_R/\sqrt{N}$, where $N$ is the recoil energy. This scale is approximately $30$~Hz, which also sets the maximum achievable instability rate. The vestigial-order phase is achievable whenever the lattice depth along the $z$ direction is large enough that the bare $E_\chi \alt 30$~Hz. Using standard data for Rb~\cite{greinernote}, we estimate that this requires a lattice depth that is $15 E_R$. This is larger than the (sub-Hz) rate at which momentum excitations decay due to spontaneous emission or contact interactions. 


\btext{In realistic experiments, cavity losses may lead to decoherence and thus destroy the VO phase. We expect that the VO phase can still be observed whenever the relaxation time to the VO phase is shorter than the decoherence time.}
\rtext{Indeed, the lifetime of the VO phase might be longer than that of the superradiant phases, since it has less photonic admixture than the superradiant phases, and thus might be more robust against photon decay; we will address this possibility in future work.}
To provide an estimate for relaxation time to the VO phase we make an estimation for the gap within the VO phase (see Appendix~\ref{sec:gap}). For the parameters $E_\chi=0.01$ and $\gamma=\gamma'=0.1$ in the units of $\xi$ (this is the same parameters as used for the Fig.~\ref{pdVO}) we obtain the the relaxation time of the order of $\tau_\text{rel}\approx 2$ sec (assuming that $\xi\approx 30$~Hz). 

\section{Vestigial order beyond two cavities}\label{multimode}

The bulk of this paper considered systems having two degenerate cavity modes, corresponding to physically separate cavities. However, there has been considerable experimental progress in realizing systems of ultracold atoms coupled to multimode optical cavities~\cite{kollar2015, kollar2016, simon_pra}. Previous theoretical work~\cite{ringcav, us, us:pra, chang_waveguide, douglas2015} argued that the self-organization transition should persist in such systems, although it might generally become first-order~\cite{us}. We now revisit this problem in light of the previous discussion, and argue that VO phases should generally be present in these geometries. 

Consider, for concreteness, the case of a transversely pumped concentric cavity, as introduced in Refs.~\cite{us, us:pra}. In this setup there is a family of cavity modes (and corresponding density-wave-like ordered states) parameterized by radial and angular quantum numbers $m, n$ (approximately enumerating the nodes in each direction) such that $m + n$ is fixed. Each such configuration has approximately the same kinetic energy per particle, on the order of a recoil energy. However, an atomic configuration that merely mixes the modes $m, n$ and $m+1, n-1$ requires very little kinetic energy, and mixes every neighboring pair of cavity modes. Thus, it softens by an amount proportional to the number of modes in the cavity. A closely analogous phenomenon occurs in the case of a photonic-crystal waveguide in which the band to which the atoms are coupled is a $p$-type band. Again, self-organization costs kinetic energy on the order of the recoil energy, whereas the VO phase costs parametrically less kinetic energy (by an amount set by the system size). Thus, the present mechanism applies with minor changes to those problems also.

\section{Conclusion}\label{conclusion}

In this work we have argued that the self-organization transition of ultracold atoms in optical cavities changes its character dramatically when the cavity has more than one degenerate mode. In particular, a vestigially ordered (VO) phase, featuring density modulations but no superradiance, emerges between the normal and superradiant phases (Fig.~\ref{pdVO}). The mechanism for vestigial order in the present context is very different from that in, e.g., high-temperature superconductors~\cite{berg_intertwined}. There, the VO phase occurs because it is more robust to disorder than its parent ordered states. Here, on the other hand, the VO phase occurs because the energetics of a long-wavelength density-wave are more favorable than those of a short-wavelength density wave. We have presented a concrete protocol for realizing the VO phase in a setup consisting of two optical cavities at an angle to one another; we estimate a maximal growth rate for the VO phase of order $30$~Hz, which should in principle make it accessible with present-day experiments. We have also argued that a many-mode geometry would further stabilize this phase. Many questions remain for future work, however, such as the nature of the fluctuations around the VO phase and the transition from the VO phase to regular superradiance---particularly in the highly multimode case---as well as the nonequilibrium dynamics of ordering~\cite{kbs}. We note that there are intriguing formal parallels between the present system and the case of bosonic mixtures near a Feshbach resonance~\cite{radzihovsky_weichman}---their ``molecular'' superfluid corresponds to the VO phase while their ``atomic'' superfluids correspond to the regular superradiant phases---although the cavity QED platform avoids some of the competing instabilities that arise near a bosonic Feshbach resonance. We also point out that VO type phase can be realized using ion chains~\footnote{T. Shi and J.I. Cirac, unpublished.}

\emph{Note added}.---While our work was being completed, a paper on a related topic appeared~\cite{lang}. The regime considered in that work (as in the existing experiments) is one in which the VO phase is not expected to be present.

\section{Acknowledgments}

The authors thank Emanuele Dalla Torre, Tobias Donner, Tilman Esslinger, Michael Fleischauer, Jonathan Keeling, Julian L\'eonard, Benjamin Lev, Andrea Morales, Giovanna Morigi, Leo Radzihovsky, and Philip Zupancic for helpful discussions. We also thank Tao Shi and Ignacio Cirac for many valuable insights into analysis of Gaussian wavefunctions. The authors acknowledge support from
Harvard-MIT CUA, NSF Grant No. DMR-1308435,
AFOSR Quantum Simulation MURI, and
AFOSR grant number FA9550-16-1-0323.

%



\begin{appendix}

\section{Mapping into the low-energy Hamiltonian}
\label{sec:map}
In this section we provide a procedure used for mapping the Hamiltonian into the low-energy sector as well as provide an exact expressions for th three-mode Hamiltonian~\eqref{eq:mft} in the main text. 

In Sec.~\ref{model} we started with the initial Hamiltonian $\mathcal{H}$~\eqref{eq1} where we separated the microscopically occupied state  $\hat \phi_0 = \sqrt{N} $ and introduced fluctuations around this state $\hat \phi_{\vec k} $. We truncated the Hamiltonian by using only leading order scattering processes (for clarity let's denote this Hamiltonian $\mathcal{H}_\text{tr}$). Then working with the quadratic part of the truncated Hamiltonian $\mathcal{H}_q$~\eqref{quadHam} in Sec.~\ref{polaritons} we introduced the polariton modes $\hat x_n$ and $\hat \pi_n$ in cavity $n$. Then the Hamiltonian $\mathcal{H}_\text{tr}$ is written using the polariton operators $\hat x_{\alpha,n}$ and $\hat \pi_{\alpha,n}$ with parameters $\lambda_{\alpha,n}$. For each cavity $n$ there are three polariton branches. Since we are interested only in the low-energy sector of the problem we discard two polariton branches and work only with the branch that softens due to the light-matter coupling. Thus, the effective Hamiltonian in the low-energy polariton picture, including the $\chi$-mode, reads

\bea\label{eq:lowenergy}
\mathcal{H}_3 & = & \sum_{n=1}^2 \left( \frac{\hat \pi_n^2}{2} + \frac{\lambda_n \hat x^2_n}{2} \right) + \frac{\hat \pi_\chi^2}{2} + \frac{E_\chi^2 \hat x_\chi^2}{2} \\
& & + \xi \hat x_\chi \hat x_1 \hat x_2+  \sum_{m,n =1}^{2}\gamma_{mn} \hat x_m^2 \hat x_n^2 + \sum_{n=1}^2 \gamma'_n \hat x_\chi^2 \hat x_n^2  \nonumber.
\eea
where the cubic non-linearity coefficient is 
\begin{eqnarray}
\xi &=& \frac{1}{2 \sqrt{2}} \mathcal{G} \sqrt{N}\sqrt{\Delta_1 \Delta_2E_{\chi}} u_1^{(a)} u_2^{(a)} \\ \nonumber
&+&  \frac{\Omega\sqrt{  E_{\chi}} }{2 \sqrt{2}} \sum_{n\neq m, \sigma=\pm} \sqrt{\Delta_n E_{m\sigma}} u_n^{(a)} u_{m\sigma}^{(\phi)} \\ \nonumber
\gamma_{nm} &=& - \frac{\Omega}{2^3 \sqrt{N}} \sum_{\sigma \sigma'=\pm} E_{n\sigma} \sqrt{\Delta_m E_{m\sigma'}} (u_{n\sigma}^{{\phi}})^2 u_m^{(2)}u_{m\sigma'}^{(\phi)} \\ \nonumber
\gamma_{n}' &=& - \frac{\Omega}{2^3 \sqrt{N}} \sum_{\sigma=\pm} E_{\chi} \sqrt{\Delta_n E_{n\sigma}}  u_m^{(2)}u_{m\sigma}^{(\phi)}
\end{eqnarray}
Recall (Sec.~\ref{polaritons}) that the coefficients $u^{(a)}, u^{(\phi)}$ denote atomic and photonic components of the $\alpha$ polaritons, and $\sigma = \pm$ refers to the upper ($+$) and lower ($-$) bare energies of the atomic modes that are coupled to cavity mode $n$.
Close to the phase transition cubic and quartic coefficients vary weakly, thus, for simplicity we can approximate the Hamiltonian ~\eqref{eq:mft} with its more simplified version when coefficients are taken at the high-symmetry line, $\Delta_1=\Delta_2$ and $E_{1\sigma} = E_{2\sigma}$. In this case the non-linearity can be described with two scalar coefficients $\gamma$ and $\gamma'$. 

\begin{figure}[tb]
\begin{center}
\includegraphics[width = 0.45\textwidth]{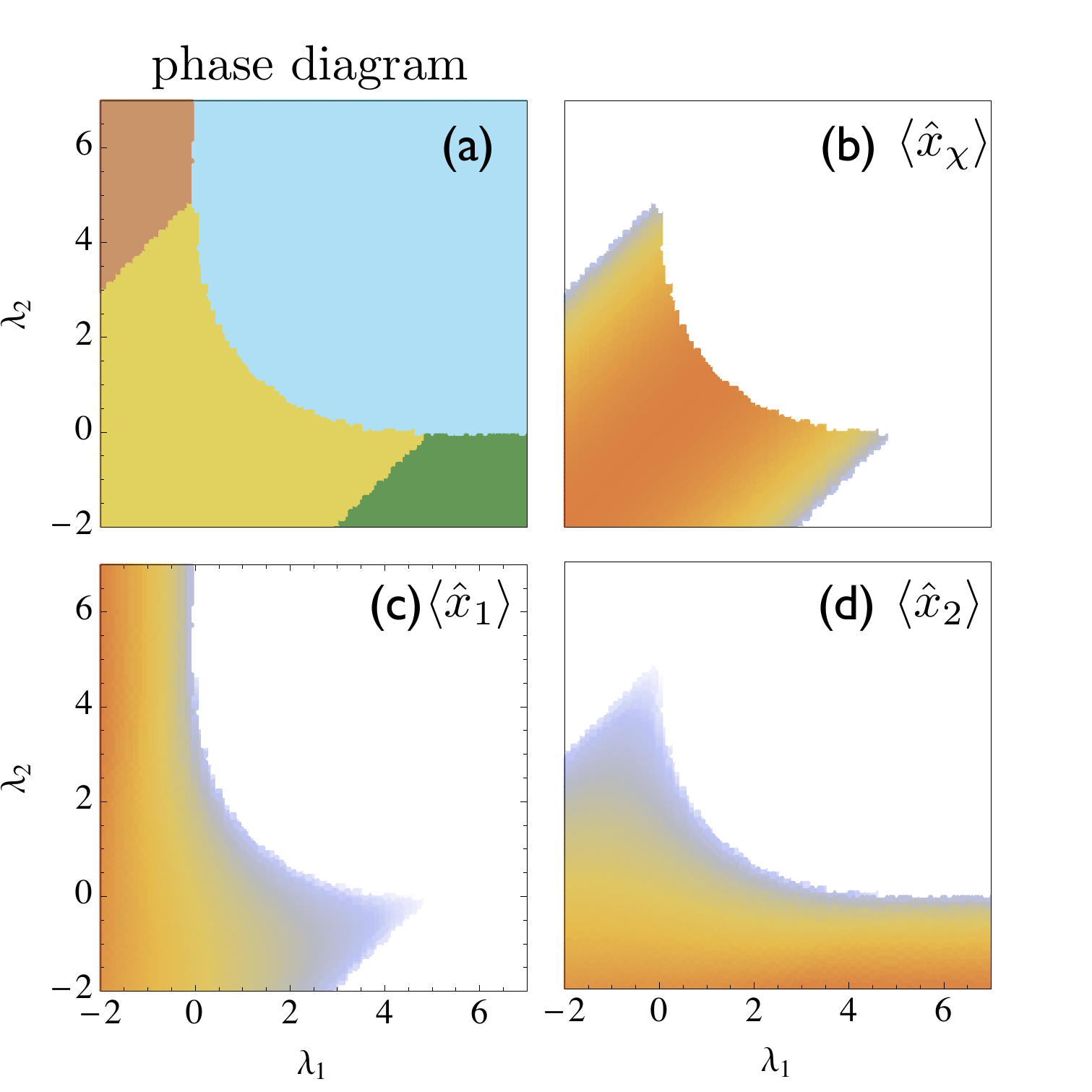}
\caption{Three-mode mean-field  (a) phase diagram and order parameters (b) $\av{\hat x_\chi}$, (c) $\av{\hat x_1}$, and (d) $\av{\hat x_2}$ as a function of the detunings of the two cavities, $\lambda_1$ and $\lambda_2$. The situation shown here is for the case when $E_\chi$ is fixed at $0.01$, in units where the strength of the cubic anisotropy is set to $\xi=1$ and the quartic anisotropy is $\gamma=0.1$. }
\label{pdmf}
\end{center}
\end{figure}

\section{Mean-field analysis of the three-mode Hamiltonian}
\btext{In this section we provide a brief analysis of the phase diagram for three mode Hamiltonian~\eqref{eq:mft}. 
As in the Sec.~\ref{sec:gaussian} we use the variational approach for describing the ground state of the system. 
At the mean-field level the wave function is given by the product of the coherent states}
\begin{equation}\label{eq:wfmf}
\ket{\Psi_\text{MF}} = e^{\sum_{i} v_i \alpha_i - \text{h.c.}} \ket 0
\end{equation}
\btext{where $v_i$ are the variational parameters. By minimizing the Hamiltonian~\eqref{eq:mft}, $\bra{\Psi_\text{MF}} \hat H \ket{\Psi_\text{MF}}$, in the state~\eqref{eq:wfmf} we obtain the phase diagram shown in Fig.~\ref{pdmf}. The transition from trivial phase to the phase with all three order parameters is a first-order phase transition, whereas the other transitions are continuous. This first-order phase transition is due to the effects of the $\chi$ mode as discussed in the main text. All other phase transitions are continuous. Note that at the level of mean-field theory, no vestigial-order phase occurs.}


\section{Gaussian approach}
We summarize the theory of Correlated Gaussian Wavefunctions (CGWs) and employ it to study the systems of correlated bosons represented by the Hamiltonian~\eqref{eq:mft}. The CGWs method is based on the variational principle using the trial wave-finction in the form~\eqref{eq:WF}. Our energy minimization procedure is based on the imginary time evolution of the system given by the density matrix evolution~\eqref{eq:densitymat}~\cite{kraus2010}.
To cast this evolution in a more practical form, we employ the equations of motion in imaginary time for the one and two operator averages (e.g. $\av{\hat O} = Tr[\hat O \hat \rho (t)]$),
\begin{eqnarray*}
\partial_\tau \av{\hat R_m} &=&  - \av{\lbrace \hat H,\hat R_m \rbrace} + 2 \av{\hat R_m} \av{\hat H}\\ \nonumber
\partial_\tau \av{\hat R_m \hat R_{m'}} &=&  - \av{\lbrace \hat H,\hat R_m R_{m'} \rbrace} + 2 \av{\hat R_m R_{m'}} \av{\hat H}\\ \nonumber
\end{eqnarray*}
where $\hat R = \left( \hat x_1 \ldots \hat x_n  , \hat \pi_1 \ldots \hat \pi_n \right)$ and $\sigma^y = \mattowbytwo{0}{-i I_3}{iI_3}{0}$. Here the average is taken over the state~\eqref{eq:WF}. To obtain the explicit form of the equations of motion the expectation values of the anticommutators should be calculated. To perform this calculation, we point out the Gaussian wavefunction allows for using the Wick's contraction, thus high-order correlation functions can be expressed through the averages of two and one operators. Equations of motion can be written in a compact form 
\begin{eqnarray}
\partial_\tau \av{\hat R_m} &=&  - \sum_{mn} \av{\hat R_m \hat R_{n}}_c h^{(1)}_{n}\\ \nonumber
\partial_\tau \av{\hat R_m \hat R_{m'}}_c &=&  \sum_{nn'}\sigma^y_{mn} h^{(2)}_{nn'} \sigma^y_{n'm'} \\ \nonumber
&-& \sum_{nn'}\av{\hat R_m \hat R_{n}}_c h^{(2)}_{nn'} \av{\hat R_n \hat R_{m'}}_c
\end{eqnarray}
where the symmetric connected part of the two operator average is defined as $\av{ \hat R_m  \hat R_{m'} }_c = \frac{1}{2}\av{\lbrace \hat R_m \hat R_{m'} \rbrace} - \av{\hat R_m} \av{\hat R_{m'}}$. Here we defined effective Hamiltonians defining the evolution in imaginary time for the one and two operator averages:
\begin{eqnarray}
 h^{(1)}_n &=& 2 \frac{\partial \av{H}}{\partial \av{R_n}}\\ \nonumber
 h^{(2)}_{nn'} &=& 4 \frac{\partial \av{H}}{\partial \av{R_n R_{n'}}_c}.
 \end{eqnarray}
 
Let us provide an example, for the most general quartic Hamiltonian in the form
\begin{eqnarray*}
\hat H &=& \sum_{ij} h_{ij} \hat R_{i} \hat R_{j}  + \sum_{ijk} \xi_{ijk} \hat R_i \hat R_j \hat R_k + \sum_{ijkl} U_{ijkl} R_{i}R_{j} R_{k} R_{l}
\end{eqnarray*}
 the average of the Hamiltonian reads:
\begin{eqnarray*}
\av{\hat H} &=& \sum_{ij} h_{ij} \av{\hat R_{i}} \av{\hat R_{j}}  + \sum_{ijk} \xi_{ijk} \av{\hat R_i} \av{\hat R_j} \av{\hat R_k}  \\ &&
+ \sum_{ijkl} U_{ijkl} \av{\hat R_{i}}\av{\hat R_{j}}\av{ \hat R_{k}}\av{ \hat R_{l}} + \\&&
\sum_{ij} h_{ij} \av{\hat R_{i} \hat R_{j}}_c  + \sum_{ijk} \tilde \xi_{ijk} \av{\hat R_i}\av{ \hat R_j \hat R_k}_c + \\ &&
\sum_{ijkl}  (\tilde U_{ijkl} +\tilde U_{klij} )\av{ \hat R_{i}}\av{\hat R_{j}} \av{\hat R_{k}\hat  R_{l}}_c + \\ &&
\sum_{ijkl}  \tilde U_{ijkl} \av{ \hat R_{i} \hat R_{j}}_c \av{\hat R_{k} \hat R_{l}}_c 
\end{eqnarray*}
where $\tilde \xi_{ijk} = \xi_{ijk} + \xi_{jik} + \xi_{jki}$ and $\tilde U_{ijkl} = U_{ijkl}+U_{ikjl}+U_{iklj}$ are the symmetrized interaction vertices. By taking the derivatives with respect to $\av{\hat R_n}$ and $\av{\hat R_n \hat R_{n'}}_c$ we obtain the state dependent Hamiltonians $h^{(1)}$ and $h^{(2)}$:
\begin{eqnarray*}
 \frac{1}{2} h^{(1)}_n &=&  \sum_j \tilde  h_{nj} \av{\hat R_j} + \sum_{jk} \tilde \xi_{njk}  \av{\hat R_j} \av{\hat R_k} +\\ &&
 \sum_{jk} \tilde \xi_{njk} \av{\hat R_j \hat R_k}_c+\\&&
 +  \sum_{jkl} U_{njkl}^p\av{\hat R_{j}}\av{ \hat R_{k}}\av{ \hat R_{l}} +\\&&
 \sum_{jkl}  \tilde U_{njkl}^p \av{R_{j}} \av{\hat R_{k} \hat R_{l}}_c 
 \\ \nonumber
 \frac{1}{4} h^{(2)}_{nn'} &=& h_{nn'} + \sum_{i} \tilde \xi_{inn'} \av{\hat R_i}  \\ &&
 + \sum_{ij} (\tilde U_{ijnn'}+\tilde U_{nn'ij}) \left(\av{\hat R_i} \av{\hat R_j} +\av{\hat R_i\hat R_j}_c \right)
 \end{eqnarray*}
 where superscript $p$ denotes the following permutation of indices $U_{mjkl}^{p} = U_{mjkl}+U_{jmkl}+U_{jkml}+U_{jklm}$ and $ \tilde U_{njkl}^p  = \tilde U_{mjkl}+\tilde U_{klmj} + \tilde U_{jmkl} + \tilde U_{kljm}$.


\section{Phase diagram: tetracritical point}
\label{sec:tetracrit}

\begin{figure}[t]
\begin{center}
\includegraphics[width = 0.45\textwidth]{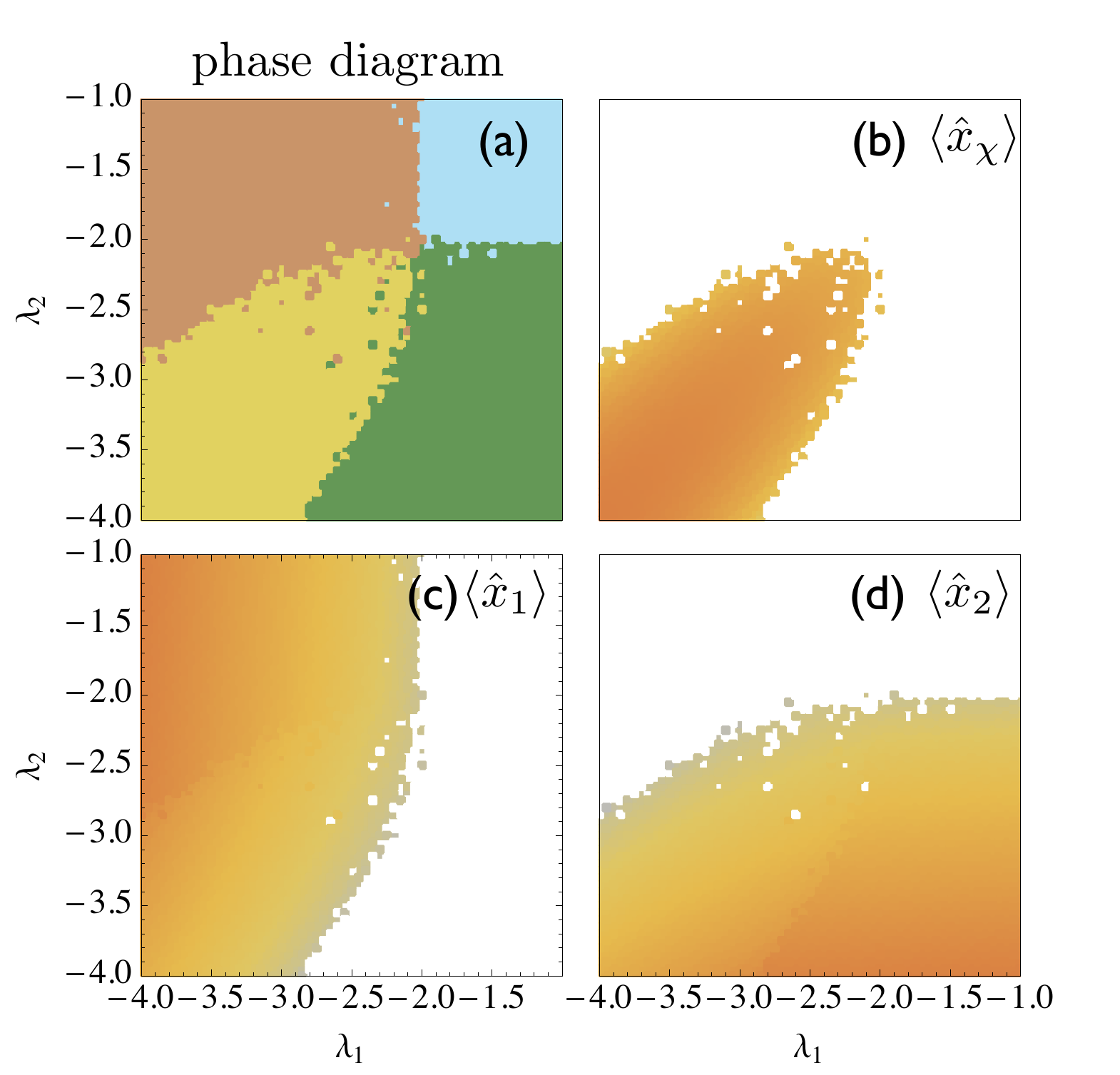}
\caption{(a) Phase diagram of the low-energy Hamiltonian~\eqref{eq:mft} and order parameters (b) $\av{\hat x_\chi}$, (c) $\av{\hat x_1}$, and (d) $\av{\hat x_2}$ as a function of the detunings of the two cavities, $\lambda_1$ and $\lambda_2$. The diagram is calculated for $E_\chi=0.01$, $\xi=1$, and $\gamma=\gamma'=0.2$. Different colors in panel (a) correspond to different phases (see the text for discussion). All phase transitions are of the second order. }
\label{pdgausnovo}
\end{center}
\end{figure}

In this section we want to point out that the vestigial order is not always present at the phase diagram even at the level when all relevant fluctuations are included. Fig.~\ref{pdgausnovo} demonstrates the phase diagram and order parameters $\av{\hat x_\chi}$, $\av{\hat x_1}$, and  $\av{\hat x_2}$ as a function of detunings for the quartic non-linearity $\gamma=0.2$ (in the units cubic anisotropy to $\xi=1$). In this case only one critical point is present in the system  (compare it with Fig.~\ref{pdVO}). This critical point separates four phases.
\begin{enumerate}
\item trivial phase $\av{\hat x_1}=\av{\hat x_2}=\av{\hat x_\chi}=0$;  
\item cavity 1 is in the superradiant phase $\av{\hat x_1}\neq 0$ and $\av{\hat x_2}=\av{\hat x_\chi}=0$; 
\item cavity 2 is in the superradiant $\av{\hat x_2}\neq 0$ and $\av{\hat x_1}=\av{\hat x_\chi}=0$;
\item all modes are macroscopically occupied  $\av{\hat x_1}\neq 0$, $\av{\hat x_2}\neq0$, and $\av{\hat x_\chi} \neq 0$. 
\end{enumerate}
There is no VO phase present on the phase diagram.

\section{Estimation for the gap inside the vestigial order phase}
\label{sec:gap}

\begin{figure}[b]
\begin{center}
\includegraphics[width = 0.45\textwidth]{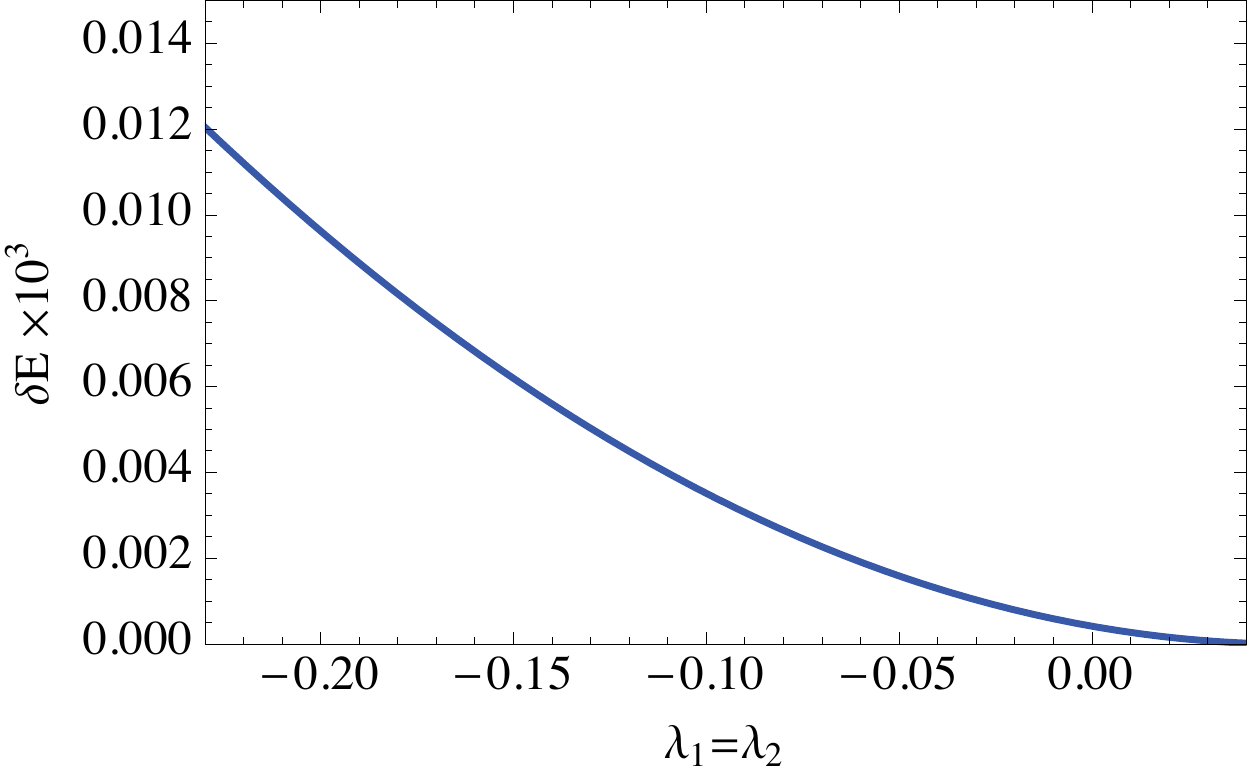}
\caption{Energy difference between the VO ground state and the trivial phase of the low-energy Hamiltonian~\eqref{eq:mft} as a function of the detunings of the two cavities, $\lambda_1$ and $\lambda_2$. The energy gap is shown for $E_\chi=0.01$, $\xi=1$, and $\gamma=\gamma'=0.1$.s}
\label{gap}
\end{center}
\end{figure}

A simple estimate for  the relaxation time to the VO phase can be made using the ground state calculation. While calculating the order parameters and correlation functions for any given point in the phase diagram we have access to the all the observables of the system calculated in the ground state. In particular, the information about the energy of the system is particularly useful. By making fit to the energy in the vicinity of the phase transition from the trivial phase to the VO phase we can estimate the energy gap between those two states inside the VO phase. In our calculations we use the quadratic fit to the energy in both phases. 

For the typical parameters used in this paper, $E_\chi=0.01$ and $\gamma=\gamma'=0.1$, we obtained an estimate for the gap equal to $\delta E = 0.015$ (all in the units of $\xi$). The dependence of the energy difference as a function of $\lambda_1$ (for $\lambda_2=\lambda_1$) is shown in Fig.~\ref{gap}. The energy difference can be increased by decreasing the ration between quartic and quibic nonlinearity $\gamma/\xi$. For instance, by decreasing $\gamma=\gamma'=0.05$ one can achieve the difference which is $\delta E = 0.07$. This result agrees with our calculation for the phase diagram for larger non-linearity $\gamma=\gamma'=0.2$ (shown in Fig.~\ref{pdgausnovo}) where the VO phase is never the lowest energy state of the system.

From this simple gap calculation we can estimate the time system need to achieve its ground state in the experiment as an inverse of the energy difference, $\tau_\text{rel} = \delta E^{-1}$. Our estimate for the aforementioned parameters are $\tau_\text{rel} \approx 2$~s and $\tau_\text{rel} \approx 0.5$~s for the quartic nonlinearity equal to $0.1$ and $0.05$ correspondingly.

\end{appendix}
\end{document}